\documentclass{cernyrep}
\usepackage{epsfig}
\usepackage{amsmath,amssymb}
\usepackage{lineno}
\usepackage[bookmarks, colorlinks=true, linktoc=page, pdftex, linkcolor=black, citecolor=black, urlcolor=blue]{hyperref}

\usepackage{fancyhdr}
\fancyhfoffset{4 mm}
\fancypagestyle{ARTTITLE}{%
\fancyhf{} 
\lhead{Proceedings of the CERN--Accelerator--School course on
{\it Introduction to Accelerator Physics}}
\lfoot{Available online at \url{https://cas.web.cern.ch/previous-schools}}
\rfoot{\thepage\hspace*{3mm}}
 
}
\newcommand\eps{\varepsilon}

\newcommand\pt{\partial}
\begin{document}
\title{Imperfections and Corrections}
\author{V. Ziemann, FREIA, Uppsala University}
\date{\today}
\begin{abstract}
  After a review of linear imperfections and their causes, we discuss
  how to model them, the diagnostic equipment needed to monitor them,
  and the correction algorithms to fix the problem they cause. We first
  address linear systems---beam lines or linear accelerators. In
  a later part we cover circular systems, such as storage rings.
\end{abstract}
\maketitle
\thispagestyle{ARTTITLE}
%
%
\section{Introduction}
When starting up a newly-built accelerator, we often find the beam not quite
where we designed it to be and its beam size is not quite what the computer
model had predicted. The reason for these discrepancies are, of course,
additional magnetic fields that affect the charged particles, which constitute
the beam. The cause of these undesired  fields are often misaligned magnets
or stray fields from adjacent components. Other reasons are intentionally
installed components that were not accounted for during the design phase and
are not in the computer model, for example, undulators in synchrotron light
sources.
\par
In the first part of these lectures we characterize the imperfections and
discuss methods of how to include them in computer models.
In the second part we discuss how the imperfections show up in linear
accelerators, how to diagnose what's wrong, and then how to correct them.
In the third part we do likewise for circular, or periodic, systems, such
as storage rings.
\par
As a prerequisite the reader should be familiar with the material from~\cite{TBD}
and~\cite{LBD}.
%
%
\section{Imperfections}
%
\begin{figure}[tb]
\begin{center}
\includegraphics[width=0.6\textwidth]{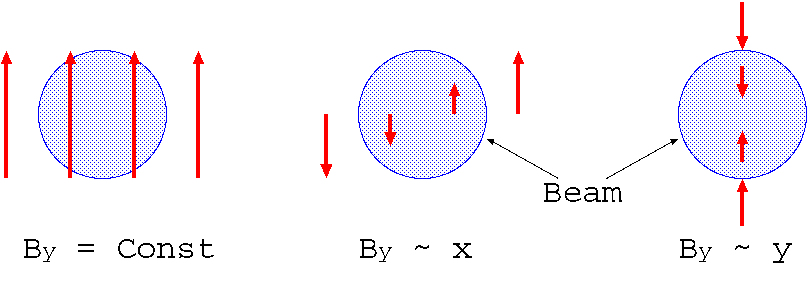}
\caption{\label{fig:tofields}Types of fields (red) that cause imperfections;
  either constant across the beam (left), with a gradient (middle) or with a
  skew-gradient (right).}
\end{center}
\end{figure}
We will predominantly deal with {\em linear} imperfections; they affect the
linear optics of the accelerator. The types of fields that cause these imperfections
are schematically shown in Figure~\ref{fig:tofields}, where the beam is depicted
as a shaded blue circle. The (transverse) field can be constant across the beam,
as shown of the left-hand side, which causes all particles to receive the same
transverse change of angle---a kick---$\Delta x'$ or $\Delta y'.$ This type of field
resembles that of a dipole corrector. A second type of field can vary linearly
across the beam, such that the particles receive a kick that is proportional
to their transverse position, as shown on the middle in Figure~\ref{fig:tofields}.
This type of field resembles that of a quadrupole. A third option is shown on the
right-hand side and resembles that of a quadrupole that is rotated by $45^o$---a
skew-quadrupole. Note that these fields correspond to the lowest-order terms of
a multipole expansion. Apart from the transverse fields, shown in
Figure~\ref{fig:tofields}, can solenoids, which may be part of high-energy
physics detectors or electron coolers, cause longitudinal fields, which we, however,
do not discuss further.
\subsection{Alignment}
Misaligned magnets are the prime sources of imperfections. The problems they cause
are mitigated by placing the magnets on alignment tables with attached pods that
are aligned to
the magnetic centers of the magnets. Surveyors then use triangulation with respect
to reference points in the tunnel to correct the positions of the magnets. The
achievable tolerances are on the order of $0.2\,$mm or, with additional effort,
somewhat better. Significantly better alignment, down to the resolution of the
beam-monitoring system, requires beam-based methods.
\par
Apart from transversely displacing magnets, the magnets can be tilted in the 
$x$-$s$--plane, where the entrance of a magnet is displaced towards one side
and the exit towards the other. And yet another misalignment is caused by
a roll angle around the direction of propagation $s$. 
\par
In the following section we discuss how to model these imperfections, which is
necessary in order to understand them and develop correction methods in later
sections.
\subsection{Modeling misalignment}
\label{sec:modmis}
Since we cannot place the magnets with infinite precision, we need to be able to
simulate their misalignment in computer codes. Let us consider one transverse
direction $x$ only. A particle with initial coordinates $\vec x_i=(x_i,x'_i)$,
passing an element, characterized by transfer matrix $\tilde R$ that is
displaced by $d_x$, is modeled by first displacing the particle by $d_x$, then
passing through the element, and finally adding the displacement $-d_x$ to the
particle coordinates. This is illustrated in the following sketch and equation.
\newline
\includegraphics[width=0.99\textwidth]{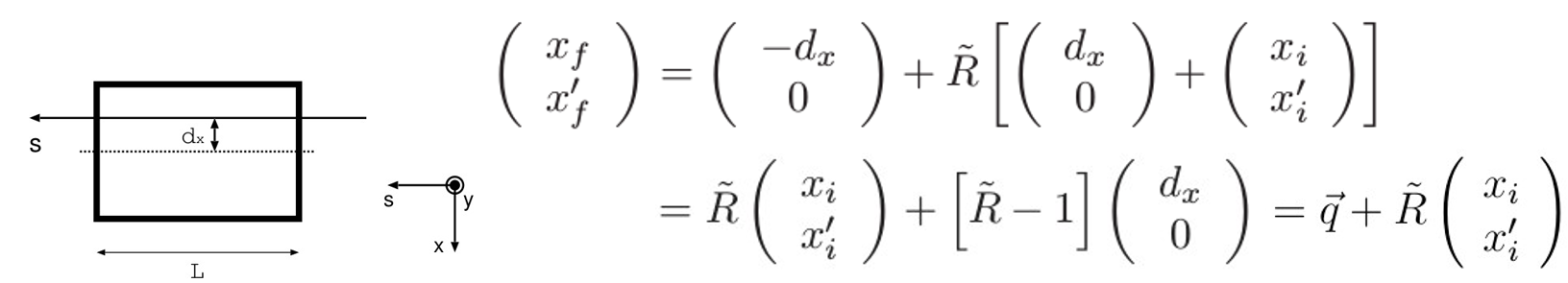}
\newline
The algebraic manipulations show that the final coordinates $\vec x_f=(x_f,x'_f)$
are given by $\vec x_f=\vec q+\tilde R \vec x_i$, which equals the un-misaligned
propagation $\tilde R \vec x_i$ and an additional term $\vec q$, which describes
an additional kick. For a thin-lens quadrupole with focal length $f$ it is easy
to show that $\vec q = (0,-d_x/f)$. We also note that the focusing of a quadrupole
is not affected by the misalignment, only the steering is, because $\vec q$ does
not depend on $\vec x_i$ and all particles receive the same kick.
\par
The effect on the beam of a magnet with length $L$ and transfer matrix $\hat R$,
tilted by $d'_x$ in the $x-s$--plane, is described by first adding $(-d_xL/2,d'_x)$
to $\vec x_i$ before passing through the magnet and finally adding  $(-d_xL/2,-d'_x)$
to the particle coordinates. Performing these step algebraically, which is left as
an exercise, shows that the result is again $\vec x_f=\vec {\hat q} + \hat R \vec x_i$,
where $\vec{\hat q}$ depends on the transfer matrix $\hat R$, the length $L$, and the
misalignment angle $d'_x$.
\par
Magnets that are rolled around the $s$--direction are modeled with the help of a
coordinate rotation in the $x-y$--plane, as shown in the following figure and
equation.
\newline
\includegraphics[width=0.99\textwidth]{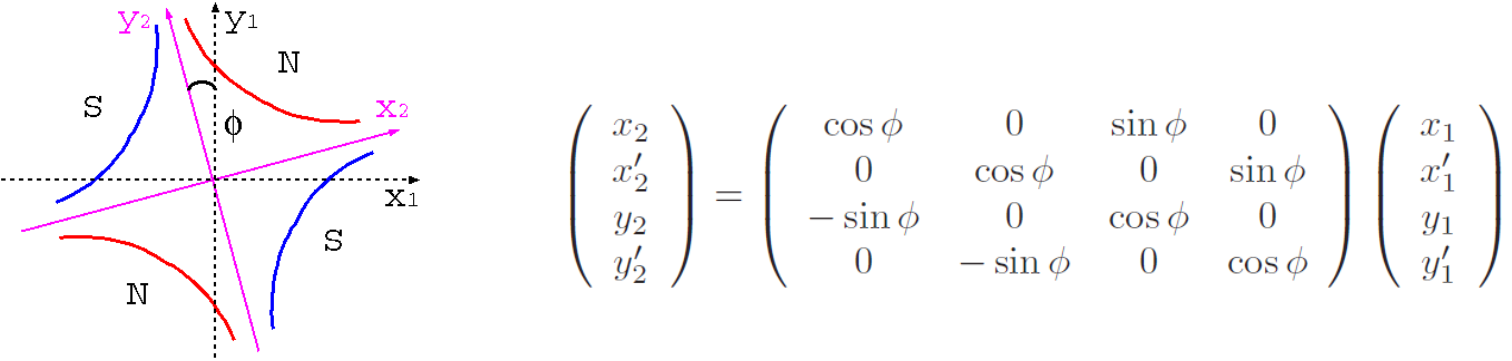}
\newline
We denote the matrix in the previous equation by $R(\phi)$. A rolled magnet can
then be described by first rotating the coordinate system with $R(\phi)$, applying
the transfer matrix of the magnet $M$, and then rotating the coordinate system back
with $R(-\phi)$. Thus, the transfer matrix of a rolled element is given by
$R(-\phi)MR(\phi)$.
\subsection{Focusing errors}
A further class of imperfections are caused by incorrectly powered quadrupoles. For
example, a focusing  quadrupole that excited too strongly, will focus the
particles to a point closer to the quadrupole. This will cause the beam (or sigma)
matrix to differ from its design values. Consequently, the beta functions will
be ``wrong'' and, in a ring, the tunes will differ from their design values.
Modeling incorrectly powered quadrupoles is accomplished by simply changing their
gradient, usually given as $k_1$, in the optics codes.
\par
Undulators and wigglers have a vertical magnetic field $B_y$ component that varies
along $s$, the direction of propagation. Maxwell's equations therefore cause the
longitudinal component $B_s$ to vary vertically, because of $\pt B_y/\pt s=
\pt B_s/\pt y$. The horizontally undulating particles therefore cross a non-zero
longitudinal field and experience a vertical force, which can be shown to be
focusing. This is a weak effect but can in some circumstances affect the
orbit and the focusing of the particles, especially when changing the field
by adjusting the gap of the undulator.
\subsection{Dispersion and Chromaticity}
Yet another class of imperfections is caused by the unavoidable spread of relative
momenta $\delta=\Delta p/p$, because the deflection of the particles is proportional
to $B/p$, thus inversely proportional to the momentum $p$. Therefore, every dipole magnet
acts like a spectrometer and separates the particles dependent on their momentum.
The position of particles is therefore to first order proportional to their
relative momentum deviation $\delta = \Delta p/p$ and given by $x=D(s) \Delta p/p$ with
the {\em dispersion function $D(s)$.} Note that the dispersion varies along the
accelerator and depends on the position $s$. A finite value of the dispersion
increases the beam size. In planar accelerators, this effect only
affects the horizontal beam size, but finite alignment tolerances can also
cause vertical dispersion to appear.

Not only the kick that the particles receive depends on their momentum, also
their focusing is affected. This momentum-dependent focusing is called
{\em chromaticity} and affects beam matrix and beta functions. In rings, also
the tunes become momentum-dependent and instead of a single value for the entire
beam, chromaticity causes a spread of tune values.
\par
We can measure dispersion and chromaticity by changing the beam energy and observing
the ensuing change in the beam position (dispersion: $D=\Delta x/(\Delta p/p)$).
In rings, we can change the frequency $f_{RF}$ system , which causes the beam to
adjust its energy to remain synchronous with the RF system and in a linear
accelerator we can change the amplitude or phase of part of the accelerator.
Optionally, we can scale all magnets by the same factor, which is equivalent to
changing the beam energy, because all observable effects are proportional to $B/p$.
\subsection{Multipoles and feed-down}
Sextupoles and other higher-order multipoles are included in accelerator lattices
in order to correct undesirable aberration. This works nicely, if they are aligned
properly. It turns out that misaligned multipoles cause additional multipoles to
appear. To quantify this effect, we remember that transverse magnetic fields are
described by the multipole expansion
\begin{equation}\label{eq:multipoles}
B_y+iB_x=B_0\sum_{m=1}^{\infty}(b_m+i a_m) \left(\frac{x+iy}{R_0}\right)^{m-1}\ ,
\end{equation}
where $B_0$ and $R_0$ are reference values and $b_m$ and $a_m$ characterize the
magnitude of the multipole component. The $b_m$ describe magnets, which only have
a vertical field component $B_y$ along the $x$--axis. They are called upright
multipoles, whereas the $a_m$ describe magnets which are rolled by $\phi=\pi/m$
and are called skew multipoles.
\par
Assuming that the magnets are short, such that they only affect the angles $x'$
and $y'$ of the particles, the kicks can be written as
\begin{equation}\label{eq:multipolekick}
  \Delta x' - i\Delta y' =\frac{(B_y+i B_x)L}{B\rho}
  = -\sum_{n=0}^{\infty} \frac{k_nL}{n!}(x+iy)^n \ .
\end{equation}
Here $L$ is the length of the magnet. It is easy to show that
$k_nL = L(\pt^n B_y/\pt x^n)_{y=0}/B\rho$ for an upright magnet, where we use $B\rho=p/e$ to express the
momentum~$p$.
\par
For a magnet with a single multipole component the kick from Equation~\ref{eq:multipolekick}
simplifies to $\Delta x' - i\Delta y' =(k_nL/n!)(x+iy)^n$ and, if the magnet is
horizontally displaced by $d_x$, the kick becomes
\begin{eqnarray}\label{eq:feeddown}
\Delta x' - i\Delta y' &=&-\frac{k_n L}{n!} (x+d_x+iy)^n\nonumber\\
&=&-\frac{k_n L}{n!} (x+iy)^n - \frac{k_n L}{n!} \sum_{k=0}^{n-1}\binom{n}{k}d_x^{n-k}(x+iy)^k\ ,
\end{eqnarray}
where the second equality derives from a binomial expansion of $(x+d_x+iy)^n$. The
first term shows that the displaced multipole still does what it was supposed to do.
But additionally all lower-order multipoles $k=0,\dots,n-1$ appear. Their magnitude
can be read off from Equation~\ref{eq:feeddown}.
\par
\begin{figure}[tb]
\begin{center}
\includegraphics[width=0.6\textwidth]{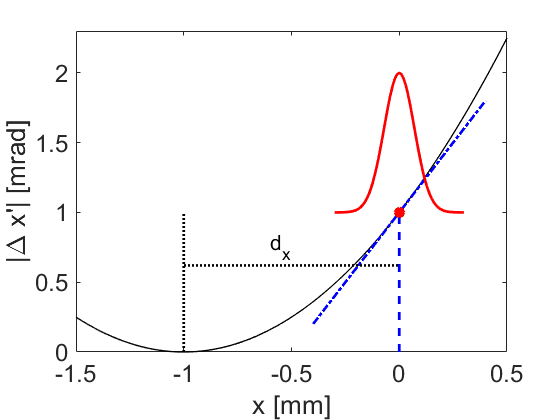}
\end{center}
\caption{\label{fig:feeddown}The black parabola denotes the kick from a sextupole
  that particles in a displaced beam, shown as the red distribution, receive. Locally,
  a constant offset, shown by the blue dashed vertical line, and a slope,
  shown by the dot-dashed blue line, are added.}
\end{figure}
These lower-order multipoles have an intuitive interpretation, which becomes apparent
when considering a horizontally displaced sextupole, whose kick is given by
\begin{equation}\label{eq:offx}
\Delta x' - i\Delta y' = -\frac{k_2L}{2} \left[(x+iy)^2 +2d_x (x+iy) + d_x^2)\right]\ . 
\end{equation}
The terms in Equation~\ref{eq:offx} are illustrated in Figure~\ref{fig:feeddown}, which
shows the absolute value of the horizontal kick $\Delta x'$ from an upright sextupole
as a function of the horizontal position $x$ as the black parabola. The displaced beam
is shown as the red Gaussian with the red dot denoting its center. The last term in
Equation~\ref{eq:offx}, proportional to $d_x^2$ describes a constant kick that affects
all particles equally. It is illustrated in Figure~\ref{fig:feeddown} by the vertical
dot-dashed line under the bunch center. The term in the middle, proportional to $2d_x(x+iy)$,
describes the slope of the parabola and illustrates that the left-hand part of the
bunch experiences a smaller kick than the right-hand part. This is just what quadrupoles
do. The first term, proportional to $(x+iy)^2$, describes the curvature of the parabola
at the position of the red dot turns out to be equal to the one in the center of the parabola.
\par
A vertically misaligned sextupole causes the particles to be kicked by
\begin{equation}\label{eq:offy}
\Delta x' - i\Delta y' =  -\frac{k_2L}{2} \left[(x+iy)^2 +2id_y (x+iy) - d_y^2)\right] \ .
\end{equation}
The term, proportional to $(x+iy)^2$, describes the sextupolar kick and the constant
term $-d_y^2$ describes a constant kick, as before. The linear term, proportional to
$d_y(x+iy)$ is now multiplied by an imaginary unit, which therefore describes a
skew-quadrupolar field. This, in turn, couples the transverse planes,
because, for example, a horizontal beam position $x$ gives rise to a vertical kick
$\Delta y'.$ In synchrotron light sources the vertical offset of the often very strong
sextupoles is one of the main causes of vertical dispersion, which spoils the vertical
emittance.
\par
So far, we discussed the imperfections of the magnets. In the next section we briefly
touch upon the imperfections of the diagnostic equipment that we will use to identify
and correct the imperfections.
\subsection{Imperfections of diagnostic components}
%
\begin{figure}[b]
\begin{center}
\includegraphics[width=0.8\textwidth]{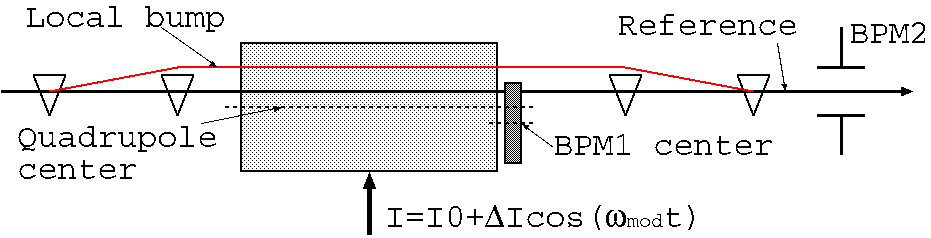}
\end{center}
\caption{\label{fig:kmod}We can determine the distance between the center of the
  quadrupole and BPM1 by K-modulating the quadrupole gradient and scanning the
  beam with a local bump across the quadrupole until the signal at the modulation
  frequency vanishes on BPM2.}
\end{figure}
Beam position monitors (BPM) are based on electronically comparing signals
from four electrodes exposed to fields that the beam generates. Tolerances in the
electronics or slight differences of their mechanical assembly can result in 
non-zero BPM readings, despite the beam being physically centered in the BPM. Tracking
down these BPM offsets is often tricky, unless the BPMs are rigidly mounted next
to a quadrupole. Figure~\ref{fig:kmod} illustrates the idea. We slightly perturb the
quadrupole with an additional sinusoidal current $\Delta I\cos(\omega_{\mathrm{mod}}t)$ and
use a local corrector bump (more on them later) to change the position of the beam
in the quadrupole. No signal with the modulating frequency $\omega_{\mathrm{mod}}$ will
show up on the second BPM2, once the beam is centered in the quadrupole. In this
state, the reading of BPM1 reveals its offset with respect to the center of the
quadrupole.
\par
Screens, inserted in the beam's path and observed by a camera, are used to
determine the beam transverse width. Fluorescent screens often have blind spots,
because they are burnt out when the beam was unintentionally parked on the screen
for extended time. The response of screens---the signal generated per nC---is often
non-linear and makes careful calibration necessary if the screens are used for
quantitative measurements. Moreover, the magnification of the optical system,
consisting of lenses, between the screen and the camera needs to be determined,
which is often accomplished by placing fiducial markers with fixed separation on
the screens. This allows to relate the pixels from the camera to the mm on the
screen. This also helps to calibrate the different scales in the horizontal and
vertical direction, if the screen is mounted at an angle.
\par
Wire scanners record the secondary emission electrons that the beam knocks
out from a wire scanned across its path. They require carefully calibrating the
position of the wire. In a {\em SEM grid} the currents from multiple wires are
read out simultaneously, which requires multiple well-balanced current amplifiers. 
\par
After having discussed the different imperfections, let us turn to linear systems,
beam lines and linear accelerator s and discuss how these effects disturb the
system and how to correct it.
%
%
\section{Linear accelerators and beam lines}
In this section we center our discussion on straight systems. The key quantities for
much of the following discussion are transfer matrix elements, especially $R_{12}$.
It describes the dependence of the position $x$, which we observe, on the cause of the
change, which is an angle $x'$. The first index in $R_{12}$ denotes the quantity we
observe, here $x$, which is the first element in the state vector of $(x,x',y,y')$.
The second index denotes what is causing the change, here it is 2, because $x'$ is
at the second place in the state vector. You might want to work out which transfer
matrix element describes the change of the vertical position $y$ due to varying the
vertical angle~$y'.$
\subsection{Transfer matrices in linear accelerators}
When calculating transfer-matrix elements in a linear accelerator, where beam energy
and momentum at the observation point---the first index---and at the ``kicking
point''---the second index---are different. Under acceleration, the longitudinal
momentum $p_s$ increases, while the transverse momentum $p_x$ remains unchanged.
This causes the beam angle $x'=p_x/p_s$ to decrease by the relativistic factor
$\beta\gamma=p_s/mc$. This effect is called {\em adiabatic damping} because it
decreases the emittance under acceleration. Moreover, $R_{12}$ scales with
$(\beta\gamma)_{kick}/(\beta\gamma)_{look}$, which we need to take into account,
when considering linear accelerators.
\subsection{Dipole errors and steering magnets}
\begin{figure}[tb]
\begin{center}
  \includegraphics[width=0.6\textwidth]{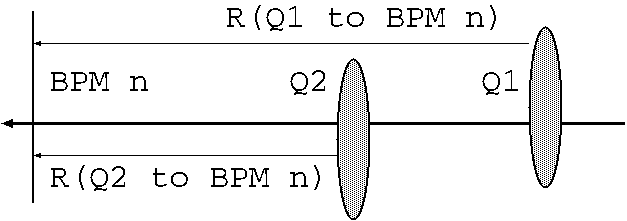}
  \hskip 5mm
  \includegraphics[width=0.35\textwidth]{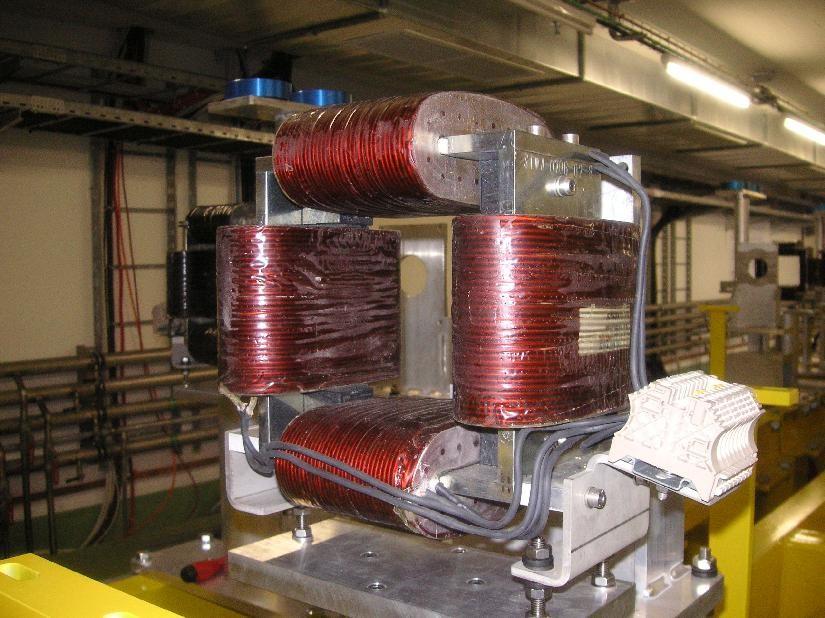}
\end{center}
\caption{\label{fig:steerer}Left: The position on the BPM(n) is determined
  my the misalignments from the quadrupoles weighted by the transfer matrix
  from the respective quadrupole to the BPM. Right: image of a steering magnet
  with coils to steer both horizontally and vertically.}
\end{figure}

Remember from Section~\ref{sec:modmis} that misalignments can be described by
applying an operator $O=\vec q + R$, consisting of a misalignment vector
$\vec q$ and the transfer matrix $R$ of the element. Here we interpret the
operation of $O$ on a state vector $\vec x$ as first applying the transfer
matrix $R$ and the adding $\vec q$ to that vector. Multiple misalignments
can therefore represented by sequentially applying operators $O_k$. The
particle coordinates at the end of the beam line, the state vector $\vec x_n$
is the given by
\begin{eqnarray}
\vec x_n &=& R_n \cdots (\vec q_{k+1} + R_{k+1})(\vec q_{k} + R_{k})\cdots (\vec q_{1} + R_{1}) \vec x_0
\nonumber\\
&=& R_n\cdots R_1 \vec x_0 + \sum_{j=1}^{n-1} (R_n\cdots R_{j+1})\vec q_j\ .
\end{eqnarray}
Inspecting the expression in the second line, we see that the final position $\vec x_n$
is given by propagating the initial state vector $\vec x_0$ with the product of all
transfer matrices from start to end, which equals what a beam line without misalignments
would cause. The sum extends over the perturbations $\vec q_j$, weighted with the
transfer matrices from the respective perturbation to the end of the beam line, which
is illustrated on the left-hand side in Figure~\ref{fig:steerer}. We can use this method
to find the influence of each misalignment vector $\vec q_j$ on the beam position at
the end $\vec x_n$.
\par
We correct these perturbation by introducing dipole-corrector magnets, such as the
one shown on the right-hand side in Figure~\ref{fig:steerer}. They apply the same
kick to all particles. The effect of a steerer on the beam is given by
\begin{equation}
\left(\begin{array}{c} x_1\\ x'_1\end{array}\right) 
= \left(\begin{array}{c} 0 \\ \theta \end{array}\right)  
+ \left(\begin{array}{c} x_0\\ x'_0\end{array}\right)  \ ,
\end{equation}
which can be cast into the same form as the misalignments, namely $\vec x_{1} =
\vec q + \tilde R \vec x_0$. We therefore can treat them like any other perturbation.
\subsection{Bumps and Knobs}
%
\begin{figure}[b]
\begin{center}
  \includegraphics[width=0.95\textwidth]{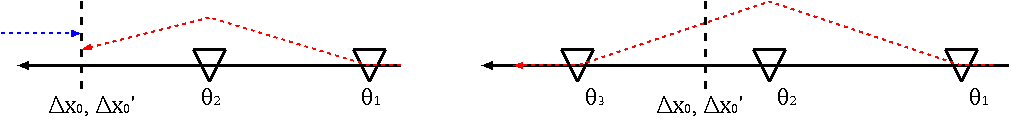}
\end{center}
\caption{\label{fig:bumps}Left: using two steerers to independently adjust position
  and angle. Right: an example of a closed three-bump, which perturbs the trajectory
  only locally.}  
\end{figure}
Often we need to combine several steering magnets to cause a well-defined change of
the beam trajectory, such as the parallel displacement to bring the blue and the red
counter-propagating beams into collision, as shown on the left-hand side in Figure~\ref{fig:bumps}.
A second example is a so-called closed bump with three steering magnets, shown on the
right-hand side in Figure~\ref{fig:bumps}, where we can adjust the position $\Delta x_0$
without perturbing the accelerator after the third corrector magnet, where the trajectory
is steered back onto the original one.
\par
We now seek linear combinations of steering-magnet excitations that achieve the
desired objective, for example, parallel displacement, which requires to adjust
$\Delta x_0$ without changing $\Delta x'_0$. We note that the first steering magnet
changes the position by $\Delta x_0=R^{01}_{12}\theta_1$ and the angle by $\Delta x'_0
=R^{01}_{22}\theta_1$. Here the subscripts denote the respective transfer-matrix element
and the first superscript denotes the objective point, here labeled ``0.'' The
second superscript denotes the
position of the steerer, here labeled ``1.'' Combining this and an equivalent equation
for the second steerer, we arrive at the following equation
\begin{equation}
\left(\begin{array}{c}\Delta x_0 \\ \Delta x'_0 \end{array}\right)
= \left(\begin{array}{cc} 
R^{01}_{12} & R^{02}_{12} \\ R^{01}_{22} & R^{02}_{22} 
\end{array}\right)
\left(\begin{array}{c}\theta_1 \\ \theta_2 \end{array}\right)\ .
\end{equation}
Since the matrix describes the response of the observables $\Delta x_0$ and $\Delta x'_0$
to a change in steerer---the actuator---it is called the {\em response matrix} for
this particular problem. Inverting the equation results in
\begin{equation}
\left(\begin{array}{c}\theta_1 \\ \theta_2 \end{array}\right)
= \left(\begin{array}{cc} 
R^{01}_{12} & R^{02}_{12} \\ R^{01}_{22} & R^{02}_{22} 
\end{array}\right)^{-1}
\left(\begin{array}{c}\Delta x_0 \\ \Delta x'_0 \end{array}\right)
\end{equation}
and gives us a way to determine the required steerer excitations $\theta_1$ and $\theta_2$
to cause a particular change in $\Delta x_0$ and $\Delta x'_0$. In particular, changing the
trajectory by $\Delta x_0$ without changing $\Delta x'_0=0$ gives us a linear combination
of steerer excitations to fulfill this objective, which is often called {\em multi-knob.}
In short, {\em the columns of the inverse of the response matrix yield the knobs} to change
one of the objective parameters.
\par
\begin{figure}[tb]
\begin{center}
  \includegraphics[width=0.95\textwidth]{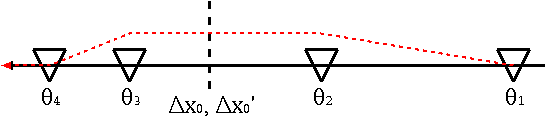}
\end{center}
\caption{\label{fig:fourbump}Using four steerers to change position and angle
  without affecting the trajectory after the last steerer, which called a four-bump.}  
\end{figure}
Let us consider the slightly more advanced example of a four-bump, which is shown in
Figure~\ref{fig:fourbump}. The objective is to independently control the position
$\Delta x_0$ and $\Delta x'_0$ at the indicated point without affecting the trajectory
after the last of the four steering magnets. To do so, we first determine the
response matrix, which is given as follows
\begin{equation}\label{eq:fourbump}
\left(\begin{array}{c}\Delta x_0\\ \Delta x'_0\\ x_f=0\\ x'_f=0 \end{array}\right)
=\left(\begin{array}{cccc}
R^{01}_{12} & R^{02}_{12} & 0 & 0 \\
R^{01}_{22} & R^{02}_{22} & 0 & 0 \\
R^{f1}_{12} & R^{f2}_{12} & R^{f3}_{12} & R^{f4}_{12}\\ 
R^{f1}_{22} & R^{f2}_{22} & R^{f3}_{22} & R^{f4}_{22}
\end{array}\right)
\left(\begin{array}{c} \theta_1 \\ \theta_2 \\ \theta_3\\ \theta_4 \end{array}\right)\ .
\end{equation}
The top left $2\times 2$ matrix equals that from the previous example. The downstream
steerers $\theta_3$ and $\theta_4$ can affect neither position $\Delta x_0$ nor angle
$\Delta x'_0$, which accounts for the $2\times 2$ matrix of zeros  in the top right corner.
The third row contains the $R_{12}$ matrix elements from each steerer to the final point
after the last steerer. Likewise the fourth row contains the $R_{22}$ elements. The vector
on the left-hand side contains the desired objectives, namely to adjust $\Delta x_0$ and
$\Delta x'_0$ while closing the bump requires both the final position $x_f$ and angle
$x'_f$ to be zero. The first and second column of the inverse response matrix respectively
are the knobs to vary position and angle independently.
%
%
\subsection{Orbit correction}
\label{sec:blorbcor}
%
\begin{figure}[tb]
\begin{center}
  \includegraphics[width=0.9\textwidth]{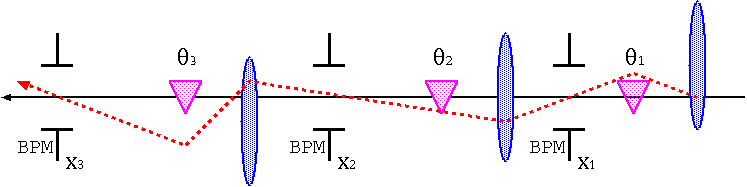}
\end{center}
\caption{\label{fig:onetoone}The transversely misaligned quadrupoles (blue) kick the beam
  (red) and the steering magnets (magenta) kick the beam back such that it passes the center
  of the BPMs. This method is called one-to-one-steering.}
\end{figure}
If the trajectory differs from some previously determined ``golden orbit'' we measure
the differences of the position recorded by the BPM and adjust steerers to zero this
difference. 
This process is called {\em orbit correction}. The simplest version is
illustrated in Figure~\ref{fig:onetoone}, where the first quadrupole is misaligned and
gives the particles a transverse kick such that the position $x_1$ on the first BPM
will be back to zero. Next we use the second steerer to correct the trajectory on the
second BPM and the third steerer to correct the third BPM. In this way we correct one
BPM at a time. This method is commonly called {\em one-to-one steering.}
\par
We can formalize the trajectory correction by introducing the  response matrix between
all BPM and steerers. Let us consider the following setup 
\newline
\includegraphics[width=0.99\textwidth]{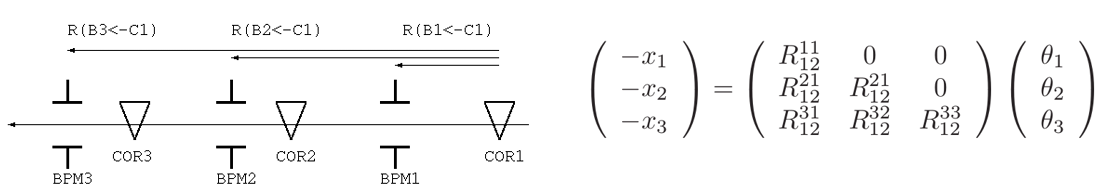}
\newline
where we show the beam line with correctors and BPM on the left-hand side and the
corresponding equation on the right-hand side. Note that the $3\times3$ response
matrix has zeros in the top right corner, because the downstream steerers cannot
affect the upstream BPMs. Otherwise the $R_{12}$ elements of the transfer-matrices
between the respective corrector and BPM appear. The vector on the left-hand side
contains the BPM readings, but with a negative sign, because we want to find corrector
values that {\em undo} the BPM readings and make them zero. Solving this equation involves
inverting the matrix, which is possible unless the response matrix is degenerate,
and gives us the steering magnet excitations $\theta_k$ to achieve this.
\par
We can calculate the response matrix with beam optics codes such as MADX~\cite{MADX}
but then the matrix is based on the model of the accelerator and may be somewhat
idealized, Moreover, neither BPM scale errors nor corrector scale errors, for
example, due to badly calibrated power supplies, are included. A second option is
therefore to determine the response matrix experimentally by first recording a
reference trajectory and observing changes of the BPM readings while changing
one steering magnet at a time.
\par
Since inverting response matrices is a very frequent task, we will look at a
number of different cases in the next section.
\subsection{Digression on linear algebra}
\label{sec:LA}
In general the systems we need to invert can be written as $-\vec x = A \vec\theta$
with the $n\times m$ response matrix $A$ for $n$ BPMs and $m$ steerers. In the
previous example we had $n=m=3$ and could simply  invert the response matrix, provided
it is non-degenerate.
\par
If we have an accelerator with more BPMs than steerers, such that $n>m$ the system of
equations, as defined by the response matrix, is over-determined and we do not have
enough steerers to correct the trajectory on all BPMs. We can, however, do our best
to minimize the rms trajectory, given by $\chi^2=\vert-\vec x-A\vec\theta\vert^2$,
which results in the well-known pseudo-inverse
\begin{equation}\label{eq:pinv}
\vec\theta = - (A^tA)^{-1}A^t\vec x\ .
\end{equation}
In very big accelerators with many BPMs and correctors, the inversion of large
matrices is numerically very sensitive, which makes using the MICADO~\cite{MICADO}
algorithm
attractive. It is based on finding the corrector that minimizes the rms orbit by
the largest amount and then implement that corrector change. In the next step the
second-most effective corrector is found and its correction applied. This process
is repeated until the trajectory is below a predetermined threshold. An added bonus
is that efficient numerical methods are used to minimize the number of computations.
\par
Finally, if the accelerator contains more steerers than BPM, the response matrix is
un\-der-de\-ter\-mined and cannot be inverted. In such cases {\em singular-value decomposition}
(SVD) is used. It decomposes $A=O\Lambda U^t$ into a diagonal matrix $\Lambda$ and two
orthogonal matrices $O$ and $U$. SVD has a very intuitive interpretation, because the
orthogonal matrices are generalized rotations and the entries on the diagonal
of $\Lambda$ are stretching factors along the rotated axes. The action of $A$ on
a vector $\vec\theta$ can thus be described by first rotating $\vec\theta$ with
$U^t$ into a coordinate system, where the axes are stretched with the factors on
the diagonal of $\Lambda$. Finally the result is rotated by $O$ into a coordinate
system which may be different from the one, where $\vec\theta$ is defined. But this
is no surprise, because the $A\vec\theta$ maps $\vec\theta$ onto a space where
the BPM positions $\vec x$ ``live.''
\par
The decomposition of $A$ now allows us to analyze where the inversion of $A$
fails, which is the case when one or several of the stretching factors on the
diagonal of $\Lambda$ are zero. These subspaces are thus projected out and cannot
be recovered. But we can still invert the matrix on the subspaces, where the
diagonals are non-zero. This entails to also project out the degenerate subspace
when calculating the inverse, which we can do by writing $"A^{-1}"=U\ "\Lambda^{-1}"\ O^t$,
where the quotes indicate that the inverse is an inverse with a twist. And twist is
to invert the diagonal matrix $\Lambda$ only where we can, namely by inverting
the entries on the diagonal where they are non-zero and project out where they
are zero. This procedure implies that wherever there is a zero on the diagonal,
we invert it by replacing $1/0$ by $0$. Finally we multiply the three inverted
matrices $U\ "\Lambda^{-1}" O^t$ and obtain $"A^{-1}"$, the inverse with a twist.
See the chapter on SVD in~\cite{NR} for a more elaborate discussion.
\par
We emphasize the usefulness of the different methods to invert matrices, because
it appears in many contexts where we can calculate the effect of control
variables on observables, such that we can calculate the response matrix
$C^{ij}=\pt$ Observable$_i/$ $\pt$ Controller$_j$. But then we need to figure out
how to set the control variables to minimize or to change the observables
by a specific amount. And that involves inverting the response matrix $C^{ij}$.
\subsection{Gradient errors and filamentation}
%
\begin{figure}[tb]
\begin{center}
  \includegraphics[width=0.8\textwidth]{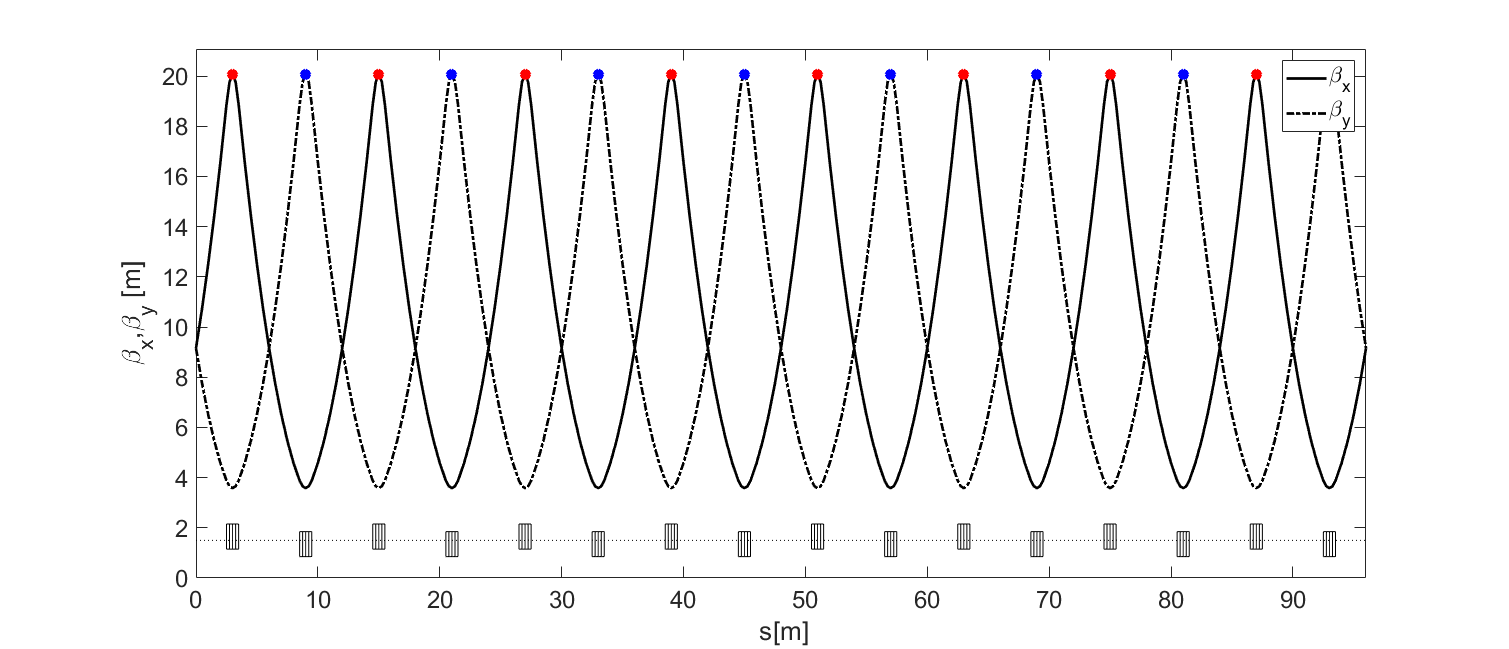}
  \includegraphics[width=0.8\textwidth]{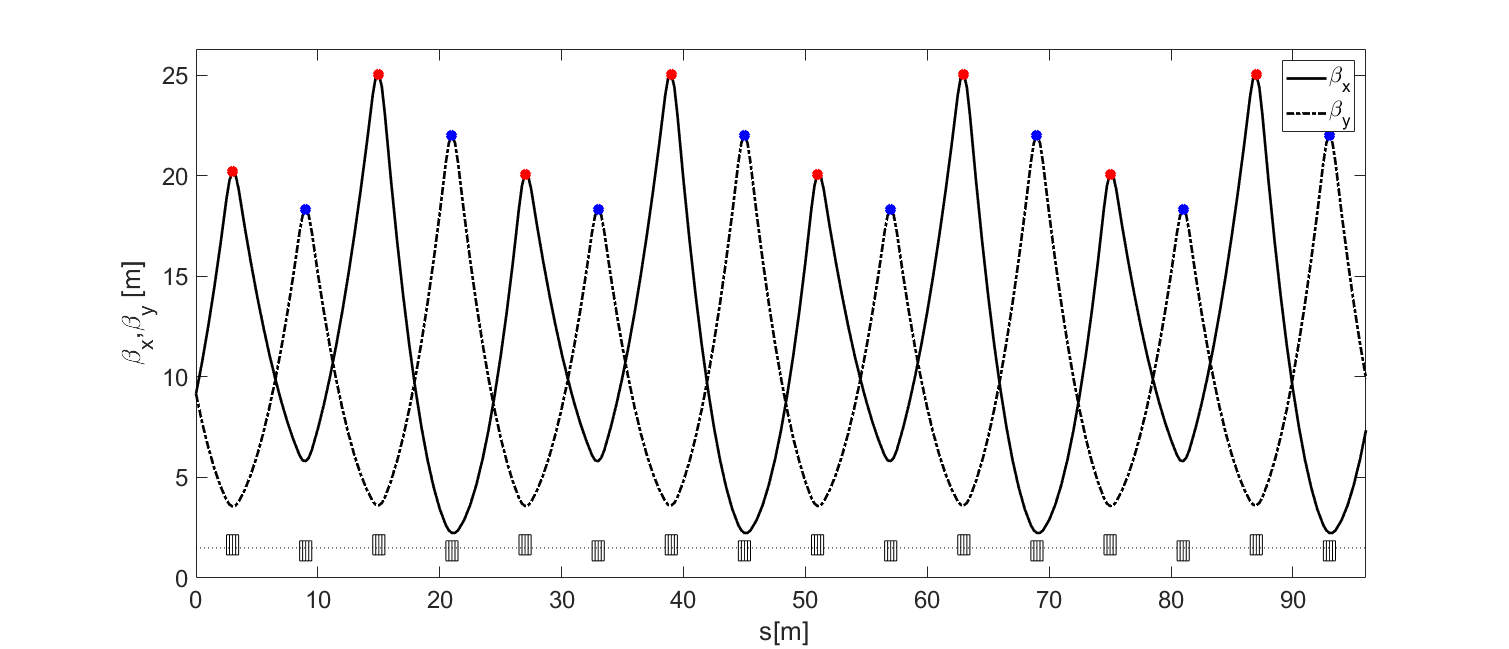}
\end{center}
\caption{\label{fig:betabeat}Top: the horizontal (solid) and vertical (dashed)
  beta functions for a beam line of eight $90^o$ FODO cells. Bottom: changing the
  first quadrupole by 10\,\% causes the beta functions to oscillate in a regular
  fashion, which is called beta beating.}
\end{figure}
Often magnetic lattices are designed to produce regular and repetitive beta
functions such as the one shown in the upper plot in Figure~\ref{fig:betabeat},
which shows $\beta_x$ (solid) and $\beta_y$ (dashed) for eight $90^o$--FODO cells.
Incorrectly powered quadrupoles or other sources of magnetic gradients,
feed down is an example, causes the beating of the beta functions, shown
on the lower plot in Figure~\ref{fig:betabeat} where the first quadrupole
has a gradient 10\,\% too low. Note the beating pattern of the red dots that
indicate the maxima of~$\beta_x$. Blue dots mark the maxima of~$\beta_y$.
It can be shown~\cite{VZA} that the beam size $\bar \sigma_x$ at a location
downstream of the error can be described by
\begin{equation}\label{eq:betabeat}
\bar\sigma_x^2 = \eps\bar\beta \left[ B_{mag} + \sqrt{B_{mag}^2-1} \cos(2\mu-\varphi)\right]\ ,
\end{equation} 
where $\eps$ is the emittance, $\bar\beta$ is the beta function at the observation
point, $\mu$ is the betatron phase advance, and $\phi$ is the starting phase.
$B_{mag}$ is called the mismatch parameter~\cite{BMAG} and is given by
\begin{equation}\label{eq:bmag}
  B_{mag}
=\frac{1}{2}\left[\left(\frac{\hat\beta}{\beta}+\frac{\beta}{\hat\beta}\right)
+\beta\hat\beta\left(\frac{\alpha}{\beta}-\frac{\hat\alpha}{\hat\beta}\right)^2\right]
\end{equation}
where $\beta$ and $\alpha$ are the unperturbed Twiss parameters and and $\hat \beta$
and $\hat\alpha$ the corresponding values with perturbation. Since $B_{mag}$ is always
larger than unity, Equation~\ref{eq:betabeat} implies that the average beam size is
increased by $B_{mag}$ and beats with amplitude $(B_{mag}^2-1)^{1/2}$ at twice the
betatron phase advance $\mu$.
\par
If we inject the beam at the end of the above transfer line into a ring, which also
constitutes a repetitive beam line, its beam size after $n$ turns is given by
\begin{equation}
\sigma_n^2 = \eps\bar\beta \left[ B_{mag} + \sqrt{B_{mag}^2-1}\ \cos(4\pi n(Q+Q'\delta)-\varphi)\right]
\end{equation}
where $Q$ is the tune of the ring and $Q'$ its chromaticity. Since the beam particles
have a distribution of relative momenta $\delta$ with width $\sigma_{\delta}$ they all
have slightly different tunes. Therefore the oscillations are no longer synchronized
and de-cohere. This mechanism is called {\em filamentation.}
We calculate the beam size $\sigma_n$ after $n$ turns by averaging over the momentum
distribution $\psi(\delta) =e^{-\delta^2/2\sigma^2_{\delta}}/\sqrt{2\pi}\sigma_{\delta}$,
which gives us
\begin{equation}
\sigma_n^2 = \eps\bar\beta \left[ B_{mag} +e^{-2(2\pi Q'\sigma_{\delta})^2n^2} 
\sqrt{B_{mag}^2-1}\ \cos(4\pi n Q-\varphi)\right]\ .
\end{equation}
We find that the beam size shows decaying oscillations towards a value that is
given by $B_{mag}$ times the unperturbed value. This can be interpreted as an
increase of the emittance by $B_{mag}$ and since $B_{mag}$ is always larger than
unity this is a very undesirable effect, especially in hadron rings without
a natural damping mechanism. Note that the decay is of type $e^{-n^2}$, which
is characteristic for de-coherence.
\subsection{Measuring the beam matrix}
\begin{figure}[tb]
\begin{center}
  \includegraphics[width=0.5\textwidth]{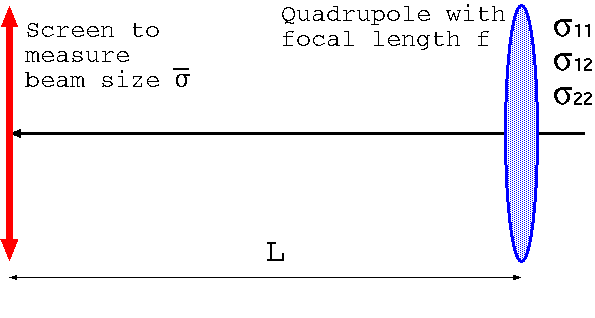}
  \hskip 5mm
  \includegraphics[width=0.4\textwidth]{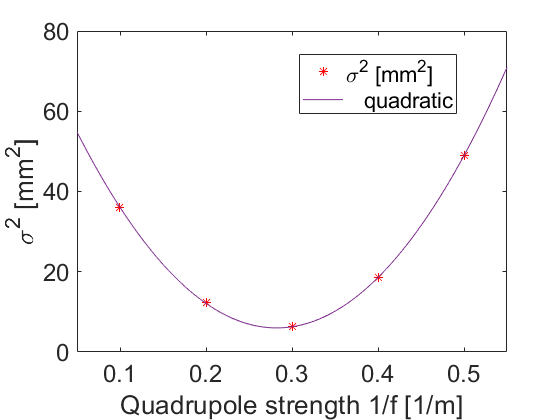}
\end{center}
\caption{\label{fig:quadscan}Left: the quadrupole changes the incoming beam with
  beam-matrix elements $\sigma_{11},\sigma_{12},$ and $\sigma_{22}$. This causes
  the beam size $\bar\sigma^2$ on a downstream screen to vary. Right:  $\bar\sigma^2$
  shows a quadratic dependence on the quadrupole excitation $1/f$.}
\end{figure}

Since small gradient errors are undesirable, yet unavoidable, we need way to
determine the beta functions experimentally before correcting them with
additional quadrupoles. A common method to measure the beam matrix and with
it the Twiss parameters and the emittance, is a {\em quadrupole scan.} It is
based on changing the quadrupole excitation and observing the changing beam
size on a screen or with a wire scanner. The setup is schematically shown on the
left-hand side in Figure~\ref{fig:quadscan}. The transfer matrix between
the quadrupole and the screen is given by
\begin{equation}\label{eq:tmq2s}
R=
\left(\begin{array}{cc}
1 & L \\ 0 & 1
\end{array}\right)
\left(\begin{array}{cc}
1 & 0 \\ -1/f & 1
\end{array}\right)
=
\left(\begin{array}{cc}
1-L/f & l \\ -1/f & 1
\end{array}\right)\ ,
\end{equation}
where $f$ is the focal length of the quadrupole and $L$ is the distance between
quadrupole and screen. If we knew the beam matrix $\sigma$ with elements $\sigma_{11},
\sigma_{12}$, and $\sigma_{22}$ we can predict the beam size on the screen $\bar\sigma$
to be
\begin{eqnarray}
\bar\sigma_x^2 &=&
R^2_{11}\sigma_{11}+ 2R_{11}R_{12} \sigma_{12} + R_{12}^2\sigma_{22}\nonumber\\
&=&(1-l/f)^2 \sigma_{11} + 2l (1-l/f) \sigma_{12} + l^2 \sigma_{22}\ ,
\end{eqnarray}
which has a quadratic dependence of $\bar\sigma^2$ on $L/f$, which is also visible on
the right-hand side in Figure~\ref{fig:quadscan}. In order to determine the $\sigma_{ij}$
from  a number of measurements, we assemble multiple---here five---measurements in a matrix 
\begin{equation}
\left(\begin{array}{c}
\bar\sigma_{x,1}^2 \\ \bar\sigma_{x,2}^2 \\ \bar\sigma_{x,3}^2 \\ \bar\sigma_{x,4}^2 \\ \bar\sigma_{x,5}^2
\end{array}\right)
=\left(\begin{array}{ccc} 
(1-L/f_1)^2 & 2L(1-L/f_1) & L^2 \\
(1-L/f_2)^2 & 2L(1-L/f_2) & L^2 \\
(1-L/f_3)^2 & 2L(1-L/f_3) & L^2 \\
(1-L/f_4)^2 & 2L(1-L/f_4) & L^2 \\
(1-L/f_5)^2 & 2L(1-L/f_5) & L^2
\end{array}\right)
\left(\begin{array}{c}\sigma_{11}\\ \sigma_{12} \\ \sigma_{22}\end{array}\right)\ .
\end{equation}
Finding $\sigma_{11},\sigma_{12}$, and $\sigma_{22}$ is now only a matter of solving this
over-determined system using the pseudo-inverse from Equation~\ref{eq:pinv}, albeit
without the minus sign. The Twiss parameters and the beta functions can be derived
from the beam matrix elements with 
\begin{equation}\label{eq:efqs}
\eps=\sqrt{\det\sigma}= \sqrt{\sigma_{11}\sigma_{22}-\sigma_{12}^2}\ ,\qquad
\beta=\frac{\sigma_{11}}{\eps}\ ,\qquad\mathrm{and}\quad
\alpha=-\frac{\sigma_{12}}{\eps}\ ,
\end{equation}
which follows from the definition of the beam matrix in terms of emittances and
Twiss parameters.
\par
Instead of using a quadrupole and a screen, we can also use several, at least three,
wire scanners in a beam line and deduce the incoming beam matrix from size measurement
on the scanners as follows
\newline
\includegraphics[width=0.9\textwidth]{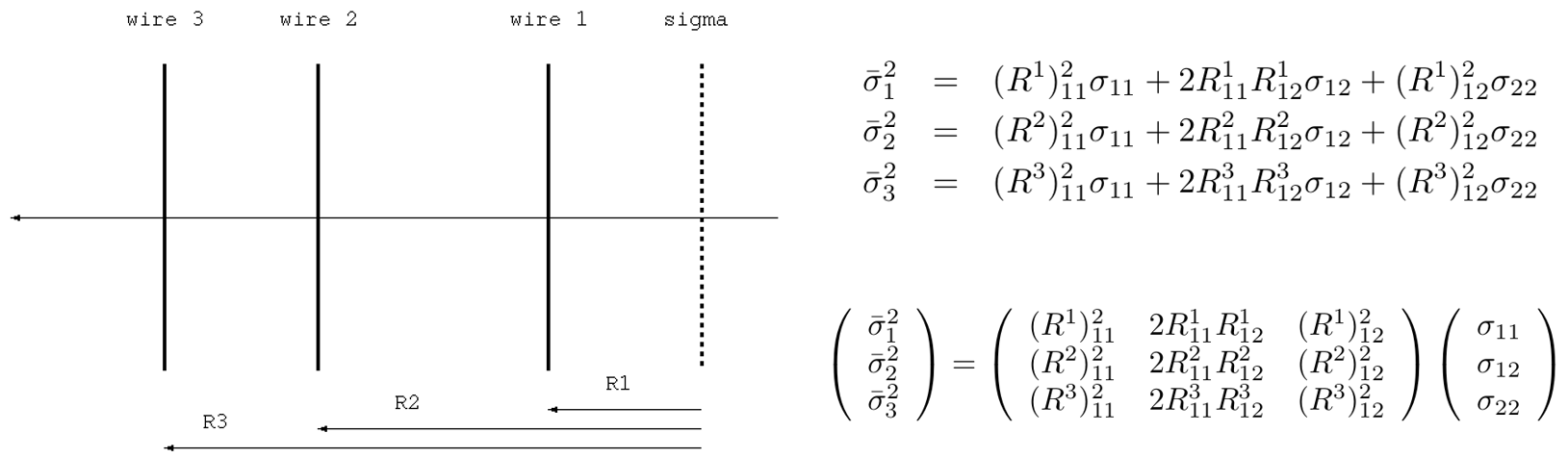}
\newline
If we know all transfer matrices, here from the reference point to the respective
wires, we can predict what we would measure, if we knew the incoming beam matrix
elements  $\sigma_{11},\sigma_{12}$, and $\sigma_{22}$. This permits us to set up
the equations shown on the right-hand side and transform them to a matrix-equation,
which we can invert with one of the methods from Section~\ref{sec:LA}.
\subsection{Correction and beta matching}
%
\begin{figure}[b]
\begin{center}
  \includegraphics[width=0.9\textwidth]{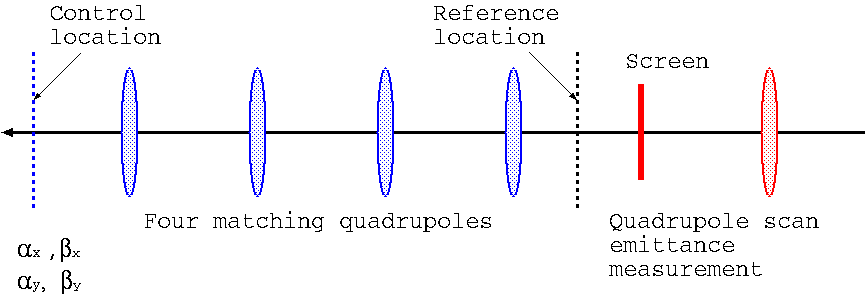}
\end{center}
\caption{\label{fig:betamatch}After determining the sigma matrix at the
  reference location in the red section, the four blue quadrupoles adjust the
  four Twiss parameters $\beta_x, \alpha_x$, $\beta_y,$ and $\alpha_y$ at
  the control location.}
\end{figure}
Using the measured beam matrix, we can use it to correct the Twiss
parameters $\beta_x, \alpha_x$, $\beta_y,$ and $\alpha_y$ at a control location, for
example, the injection point into a ring, in order to prevent emittance growth due
to filamentation. Figure~\ref{fig:betamatch} illustrates the setup with a sigma
measuring section, shown in red, and four independently powered quadrupoles, shown
in blue that can independently adjust the Twiss parameters at the injection, or
control, point. We point out that $\beta_x,\alpha_x,\beta_y,$ and $\alpha_y$ at
the control location have a non-linear dependence on the quadrupole excitation.
Finding these excitations, based on the knowledge of the incoming beam matrix
at the reference location, to set the Twiss parameters to their design values
involves non-linear optimization, commonly called {\em matching.} Beam optics
codes, such as MADX~\cite{MADX}, provide functions to specify Twiss
parameters at the start and end of a section and then suitably adjust the
excitations of the quadrupoles to match the specified boundary conditions.
\par
If the discrepancy of the actually measured Twiss parameters at the reference
position is not too far from their design values, we can calculate a linearized
response matrix of the dependence of $\beta_x,\alpha_x,\beta_y,$ and $\alpha_y$
on the four quadrupole excitations and calculate {\em knobs} to independently adjust
one of the four Twiss parameters without affecting the others. These knobs are
thus suitable linear combinations of quadrupole excitation patterns to correct
one parameter at a time.
\par
An even simpler example is a so-called {\em waist knob} that uses two
quadrupoles near an interaction point. It uses two quadrupoles to independently
control $\alpha_x$ and $\alpha_y$, or equivalently the longitudinal position of
the focal point---the waist. As in the previous paragraph, the knob is constructed
from the response matrix that relates  $\alpha_x$ and $\alpha_y$ to small changes
of the two quadrupole excitations. In this case, the incoming beam matrix is
assumed to have design values.
\subsection{Skew-gradient errors}
In accelerators with very flat beams, having $\eps_y\ll\eps_x$, skew quadrupoles
couple the large amplitude horizontal oscillations into the vertical plane and
spoil the small vertical emittance. In order to quantify this effect we consider
the effect of an additional thin skew-quadrupole with transfer matrix
\begin{equation}\label{eq:skewq}
S=\left(\begin{array}{cccc}
    1 & 0 &  0  & 0 \\
    0 & 1 & 1/f & 0\\
    0 & 0 &  1  & 0\\
   1/f& 0 &  0  & 1
\end{array}\right)
\end{equation}
on the vertical emittance of an initially uncouple beam matrix. After the skew
qua\-dru\-pole, the vertical lower-right $2\times2$ part of the beam matrix is
\begin{equation}
\left(\begin{array}{cc}
\hat\sigma_{33} & \hat\sigma_{34} \\ \hat\sigma_{34} & \hat\sigma_{44}
\end{array}\right)
= \left(\begin{array}{cc}
\sigma_{33} & \sigma_{34} \\ \sigma_{34} & \sigma_{44} + \sigma_{11}/f^2
        \end{array}\right)
\end{equation}
and its (projected) emittance $\hat\eps_y$ is given by the determinant
\begin{equation}\label{eq:coupinc}
\hat\eps_y^2 = \eps_y^2 + \frac{\sigma_{11}\sigma_{33}}{f^2}
= \eps_y^2 \left( 1 + \frac{\eps_x}{\eps_y}\frac{\beta_x\beta_y}{f^2}\right) \ .
\end{equation}
We observe that the vertical emittance $\hat\eps_y$ increases with
$\eps_x/\eps_y\gg 1$ and with $\beta_x\beta_y/f^2$, such that a large emittance
ratio is particularly detrimental; as are large beta functions $\beta_x$ and
$\beta_y$ at the location of the skew quadrupole.
%
%
\section{Circular accelerators}
In rings the beam ``bites its tail;'' it has to satisfy periodic
boundary conditions. This poses additional constraints on the motion. We first
address the consequence of a dipole error on the closed orbit in a ring.
\subsection{Dipole errors}
\label{sec:diperr}
%
\begin{figure}[tb]
\begin{center}
  \includegraphics[width=0.5\textwidth]{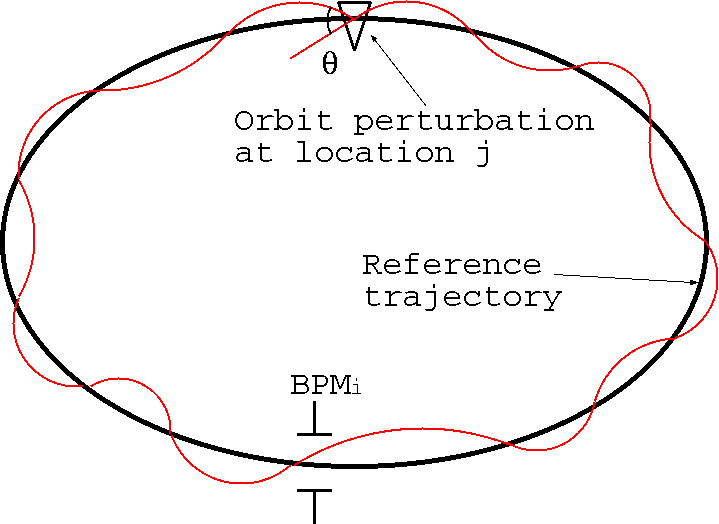}
\end{center}
\caption{\label{fig:co}Closed orbit (red), perturbed by a small dipole.}
\end{figure}
We consider a dipole error, represented as the small wedge in Figure~\ref{fig:co}.
It causes a perturbation of the closed orbit, which is shown as the red line
oscillating around the unperturbed orbit. We characterize the perturbing kick
by the vector $\vec q=(0,\theta,0,0)$, here for a horizontal kick $\theta$,
and the one-turn transfer matrix that starts at the location of the perturbation
by the $4\times 4$--matrix $R^{jj}$. The perturbed closed orbit $\vec x_j$,
immediately after the perturbation, is given by $\vec x_j = R^{jj} \vec x_j
+ \vec q_j$, which requires $\vec x_j$ to reproduce after one turn. Solving 
for $\vec x_j$ yields $\vec x_j = (1-R^{jj})^{-1} \vec q_j$ and propagating
$\vec x_j$ to the location of a BPM$_i$ with the $4\times 4$ transfer
matrix $R^{ij}$, results in the response of the BPM$_i$ to the kick at
location $j$
\begin{equation}\label{eq:clobi}
\vec x_i = R^{ij} \vec x_j =  R^{ij} (1-R^{jj})^{-1} \vec q_j = C^{ij} \vec q_j \ .
\end{equation}
Here $C^{ij}=R^{ij}(1-R^{jj})^{-1}$ is a $4\times4$--matrix that describes the
response of the closed orbit at BPM$_i$ to a perturbation at a location
labeled $j$. In this sense it takes the role that the transfer matrix has
in a beam line, but has the closed orbit constraint built in through
the factor $(1-R^{jj})^{-1}$.
\par
It can be shown that in uncoupled rings $C^{ij}_{12}$ can be written with
the help of the Twiss parameters and the phase advance between the two
locations as
\begin{equation}\label{eq:respclas}
  C^{ij}_{12} = \frac{\sqrt{\beta_i\beta_j}}{2\sin(\pi Q)}\cos(\mu_{ij}-\pi Q)\ ,
\end{equation}
where $Q$ is the tune of the ring. Note that the expression diverges at integer
values of the tune, because the sine in the denominator becomes zero.
\par
A horizontal kick will cause the closed orbit to become slightly longer; it
increases the circumference $C$ by $\Delta C=D_j\theta$ compared to the
unperturbed orbit. Here $D_j$ is the dispersion
at the location of the perturbation. In the presence of a radio-frequency (RF)
system, the beam therefore has to adjust its relative momentum by $\delta=-D_j\theta/\eta C$
to remain synchronous with the RF. This small change of momentum will
show up on BPM$_i$ as an additional displacement of the orbit by $D_i\delta,$
where $D_i$ is the dispersion at BPM$_i$, such that the response coefficient
$C^{ij}_{12}$ that includes this effect is given by
\begin{equation}\label{eq:respcof2D}
C^{ij}_{12}=\left[\frac{\sqrt{\beta_i\beta_j}}{2\sin(\pi Q)}\cos(\mu_{ij}-\pi Q) 
- \frac{D_i D_j}{\eta C}\right]\ ,
\end{equation}
where $\eta=\alpha-1/\gamma^2$ is the phase-slip factor and $\alpha$ is the
momentum compaction factor. This additional factor in $C^{ij}_{12}$ is often
neglected, but plays a role in small rings that ramp their energy, thus changing
the relativistic factor $\gamma$ to become close or even equal to $1/\sqrt{\alpha},$
a condition called {\em transition.}
\subsection{Quadrupole alignment tolerances}
The alignment tolerances for the quadrupoles can be specified by calculating the
rms orbit displacement caused by quadrupoles, transversely displaced by $d_j$, which
kick the beam by $\theta_j=d_j/f$, where $f$ is the focal length of the quadrupole.
If we assume that the displacements $d_j$ are independent, have zero mean and rms
value $\sigma_d$, we can specify their statistics by $\langle d_j\rangle=0$ and
$\langle d_jd_k\rangle =\sigma_d^2 \delta_{jk}$, where $\delta_{jk}$ is unity for
$j=k$ and zero otherwise. Using the response coefficients from Equation~\ref{eq:respclas}
we find the rms orbit displacement from summing over all quadrupoles and averaging
over the random distribution of displacements with the following result
\begin{eqnarray}
\langle x_i^2 \rangle &=& \langle
\left[\sum_j\frac{\sqrt{\beta_i\beta_j}}{2\sin\pi Q} \cos(\mu_{ij}-\pi Q) \frac{d_j}{f_j}\right]
\left[\sum_k\frac{\sqrt{\beta_i\beta_k}}{2\sin\pi Q} \cos(\mu_{ik}-\pi Q) \frac{d_k}{f_k}\right]
\rangle\nonumber\\
&=& \sum_j \frac{\beta_i\beta_j}{(2\sin\pi Q)^2}\cos^2(\mu_{ij}-\pi Q) \frac{\sigma_d^2}{f_j^2}\ ,
\end{eqnarray}
where we assumed that he phases are evenly distributed, such that we can use $\langle\cos^2\rangle
\to 1/2$. Furthermore, introducing the average beta function $\bar\beta$ at the $N_q$ quadrupole
locations and the average focal length $\bar f$, we finally arrive at
\begin{equation}
\sqrt{\langle x_i^2 \rangle}
\approx \sqrt{N_q}\frac{\bar\beta/\bar f}{2 \sqrt{2}\sin\pi Q}\sigma_d\ ,
\end{equation}
where we see that  very large rings with a large number of quadrupoles $N_q$ will
cause large rms orbit deviations, unless very tight alignment tolerances $\sigma_d$
are enforced.
\subsection{Orbit correction}
Correcting the orbit in a ring is based on calculating---or measuring---the response
matrix of how the BPM positions change as a consequence of changing steering magnets.
Since the steerers change the closed orbit, the readings of {\em all} BPM will be
affected as described by the response coefficients in $C^{ij}=R^{ij}(1-R^{jj})^{-1}$
between BPM$_i$ and corrector~$j$. For the horizontal plane, we use the $12$--elements,
such that the full response matrix is
\begin{equation}\label{eq:rorbcor} 
\left(\begin{array}{c}-x_1\\ -x_2\\ \vdots \\ -x_n \end{array}\right)
= \left(\begin{array}{cccc}
C_{12}^{11} & C_{12}^{12} & \dots & C_{12}^{1m} \\ 
C_{12}^{21} & C_{12}^{22} & \dots & C_{12}^{2m} \\ 
   \vdots   &  \vdots     & \ddots& \vdots\\
C_{12}^{n1} & C_{12}^{n2} & \dots & C_{12}^{nm} 
\end{array}\right)
\left(\begin{array}{c}\theta_1\\ \theta_2\\ \vdots \\ \theta_m \end{array}\right)\ ,
\end{equation}
which is analogous to the response matrix discussed in section~\ref{sec:blorbcor}. Keep
in mind that the superscripts label the respective BPM and steerer, whereas the subscripts
label the matrix element of the matrix $C^{ij}$. Finding
the steerer excitations $\theta_j$ that zero the orbit $x_i$ thus involves inverting the
matrix using one of the methods discussed in Section~\ref{sec:LA}.
\par
%
%
Physically, the steering magnets used for orbit correction are dipole magnets and they
also generate dispersion. Since we normally do not want to generate additional dispersion,
we include the effect in the response matrix through the dispersion-response coefficient
$S^{ij}_{12}=\pt D_i/\pt\theta_j,$ where $D_i$ is the dispersion at BPM$_i$ that can be
either calculated from the model or
measured by changing the RF frequency, which causes beam momentum to change, as already
discussed near the end of Section~\ref{sec:diperr}. In order to correct the orbit, while
minimizing the generated dispersion, which is called {\em dispersion-free steering},
we use the following augmented response matrix
\begin{equation}\label{eq:dfs} 
\left(\begin{array}{c} \vdots \\ -x_i\\ \vdots \\ -D_i\\ \vdots\end{array}\right)
= \left(\begin{array}{cccc}
\vdots & \vdots &       & \vdots\\
C_{12}^{i1} & C_{12}^{i2} & \dots & C_{12}^{im} \\ 
   \vdots   &  \vdots     &       & \vdots\\
S_{12}^{i1} & S_{12}^{i2} & \dots & S_{12}^{im} \\
\vdots & \vdots &       & \vdots
\end{array}\right)
\left(\begin{array}{c}\theta_1 \\ \theta_2\\ \vdots\\ \theta_m \end{array}\right)
\end{equation}
in which the dispersion-response coefficients $S^{ij}$ are added below the matrix from
Equation~\ref{eq:rorbcor}.
\subsection{Gradient errors}
A gradient error in a ring with tune $Q=\mu/2\pi$ can be modeled by adding a thin-lens
quadrupole at the position in the ring where the error is located, here assumed to
have Twiss parameters $\alpha$ and $\beta$. The perturbed full-turn matrix thus can be
calculated by evaluating
\begin{eqnarray}\label{eq:RQR}
R_Q R&=&\left(\begin{array}{cc} 1 & 0 \\ -1/f & 1 \end{array}\right)
\left(\begin{array}{cc}
\cos\mu +\alpha\sin\mu & \beta\sin\mu \\
-\frac{1+\alpha^2}{\beta}\sin\mu & \cos\mu -\alpha\sin\mu\\
\end{array}\right)\\
&=&\left(\begin{array}{cc}\cos\mu +\alpha\sin\mu & \beta\sin\mu\\
-(\cos\mu+\alpha\sin\mu)/f + \gamma\sin\mu & \cos\mu -\alpha\sin\mu -(\beta/f)\sin\mu
         \end{array}\right)\ .
       \nonumber
\end{eqnarray}
The perturbed tune $Q+\Delta Q$ is determined by the sum of the diagonal elements 
\begin{equation}\label{eq:ccts}
2\cos(2\pi (Q + \Delta Q)) = 2\cos(2\pi Q) - \frac{\beta}{f}\sin(2\pi Q)\ .
\end{equation}
Assuming that $\Delta Q$ is small, expanding the left-hand side to first order
gives us an approximate equation for the tune-shift $\Delta Q$, given by
$\Delta Q \approx {\beta}/{4\pi f}$, an equation that is of great practical use
as we shall see.
\par
Not only the tune, but also the beta functions change as a consequence
of a gradient error. From the $12$--element of the transfer matrix in
Equation~\ref{eq:RQR} we see that the perturbed beta function $\bar\beta$
is given by $\bar\beta\sin(2\pi(Q+\Delta Q)=\beta\sin(2\pi Q)$, or
\begin{equation}
\bar\beta=\frac{\beta\sin(2\pi Q)}{\sin(2\pi(Q+\Delta Q))}
\approx\beta\left[1+2\pi\Delta Q\cot(2\pi Q)\right] \ ,
\end{equation}
where we see that $\bar\beta$ diverges at half-integer values of the tune $Q$.
\par
Actually, a region around the half-integer tune values does not permit stable
oscillations, because Equation~\ref{eq:ccts} requires to calculate an inverse cosine
of a quantity that has magnitude larger than unity. The range of tune
values $Q$ for which 
\begin{equation}
\left\vert\cos(2\pi Q)-2\pi\Delta Q\sin(2\pi Q)\right\vert > 1
\end{equation}
exceeds unity, defines the half-integer {\em stop bands}, which depend on the
magnitude of the gradient perturbation, as quantified by $\Delta Q=\beta/4\pi f$.
\subsection{Measuring and correcting the tune and beta functions}
%
\begin{figure}[tb]
\begin{center}
  \includegraphics[width=0.9\textwidth]{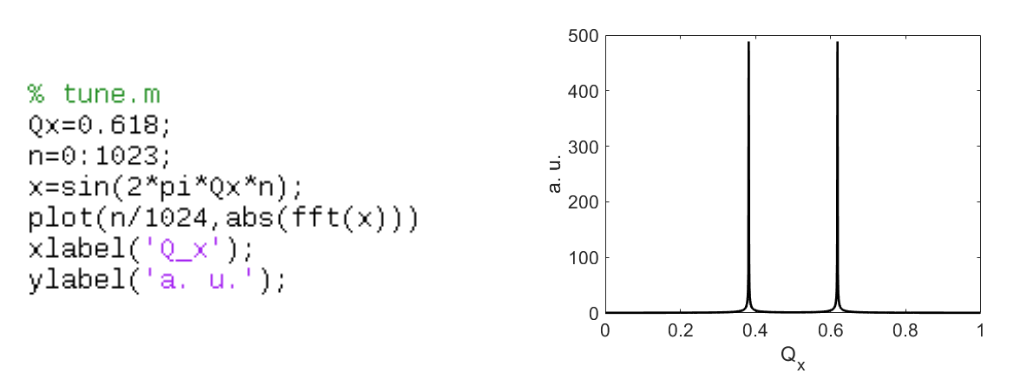}
\end{center}
\caption{\label{fig:tune}MATLAB code to produce and analyze turn-by-turn
  positions (left) and the plot showing the tune line and its alias.}
\end{figure}
The simplest way to measure the tune is to give the beam a small kick and observe
the position signal from a BPM on a spectrum analyzer, which reveals the betatron
sidebands of the revolution harmonics. Since many modern BPM provide turn-by-turn
position information, Fourier-transforming this data yields the tunes directly.
Figure~\ref{fig:tune} illustrates the process. The code on the left-hand side
defines the tune $Q_x$ and then tracks the particle for 1024 turns, before
plotting the absolute value of the FFT. Note that the initially chosen tune is
above $1/2$ such that both the original tune and its alias $1-Q_x$ appear.
This ambiguity can be resolved by slightly increasing the excitation of a
horizontally focusing quadrupole and observing the line below the half-integer.
If it moves to the right, the tune is below that half-integer, if it moves
to a lower value, the ``real'' tune is above the half-integer.
\par
Once we can measure the tunes, we can also correct it with two suitably
chosen quadrupoles. Since the quadrupoles affect both planes, the horizontal and
vertical tunes will change with the focal length $f_1$ of the first quadrupole
according to $\Delta Q_x = {\beta_{1x}}/{4\pi f_1}$ and
$\Delta Q_y =-{\beta_{1y}}/{4\pi f_1}$.
Using a second quadrupole with focal length $f_2$ their combined effect on
the tunes is given by
\begin{equation}
\Delta Q_x = \frac{\beta_{1x}}{4\pi f_1} + \frac{\beta_{2x}}{4\pi f_2}
\qquad\mathrm{and}\qquad
\Delta Q_y =-\frac{\beta_{1y}}{4\pi f_1} - \frac{\beta_{2y}}{4\pi f_2}\ .
\end{equation}
Assembling these equations into a matrix-valued equation
\begin{equation}\label{eq:tunefix}
\left(\begin{array}{c}\Delta Q_x \\ \Delta Q_y\end{array}\right)
= \frac{1}{4\pi}
\left(\begin{array}{rr}
\beta_{1x} & \beta_{2x}\\
-\beta_{1y} & -\beta_{2y}
\end{array}\right)
\left(\begin{array}{cc} 1/f_1 \\ 1/f_2 \end{array}\right)
\end{equation}
makes it obvious that the inverse of the matrix gives the excitations of the
quadrupoles that will change the tunes by $\Delta Q_x$ and $\Delta Q_y$,
respectively.
\par
Note that changing the excitation of a quadrupole by a small amount, characterized by
a small additional thin-lens quadrupole with focal length $f$ will cause a tune
shift by $\Delta Q=\beta/4\pi f$, which is proportional to the beta function at the
location of the quadrupole, thus providing a measurement of the beta function.
Often this is, however, difficult to implement, because multiple quadrupoles are
powered in series by the same power supply.
\subsection{Model calibration, LOCO}
An elaborate method to determine the differences of the accelerator in the tunnel to
the computer model is based on comparing the response coefficients $\hat C^{ij}$
obtained from measuring orbit changes as a consequence of changing the excitation
of steerers one at a time to the response coefficients $C^{ij}$ from the model.
We express the measured coefficients as the first-order Taylor expansion of the
model coefficients in the gradients~$g_k$ of the quadrupoles
\begin{equation}\label{eq:ormidea}
\hat C^{ij} = C^{ij} + \sum_k \frac{\pt C^{ij}}{\pt g_k} \Delta g_k \ ,
\end{equation}
where the derivatives ${\pt C^{ij}}/{\pt g_k}$ are calculated from the model. Note that
there are $2 N_{bpm} N_{cor}$ response coefficients in the two planes, which is normally
a very large number to determine the $N_{quad}$ gradients, which is a much smaller
number. The fit is therefore vastly over-determined. 
\par
It is straightforward to include additional parameters, such as the BPM scale errors
$\Delta x^i$ and the corrector scale errors $\Delta y^j$ which turn the equation
into
\begin{equation}\label{eq:loco}
\hat C^{ij} = C^{ij} + \sum_k \frac{\pt C^{ij}}{\pt g_k} \Delta g_k + C^{ij}\Delta x^i - C^{ij}\Delta y^j\ .
\end{equation}
This allows to reduce many systematic errors from the measurement system and even
adding further parameters is possible. These methods
were first used in SPEAR~\cite{CALIF} and later refined at NSLS~\cite{LOCO} with
remarkable success. Today, most synchrotron light sources use response-matrix based
method to debug their accelerator optics.
\subsection{Coupling and its correction}
\begin{figure}[tb]
\begin{center}
  \includegraphics[width=0.9\textwidth]{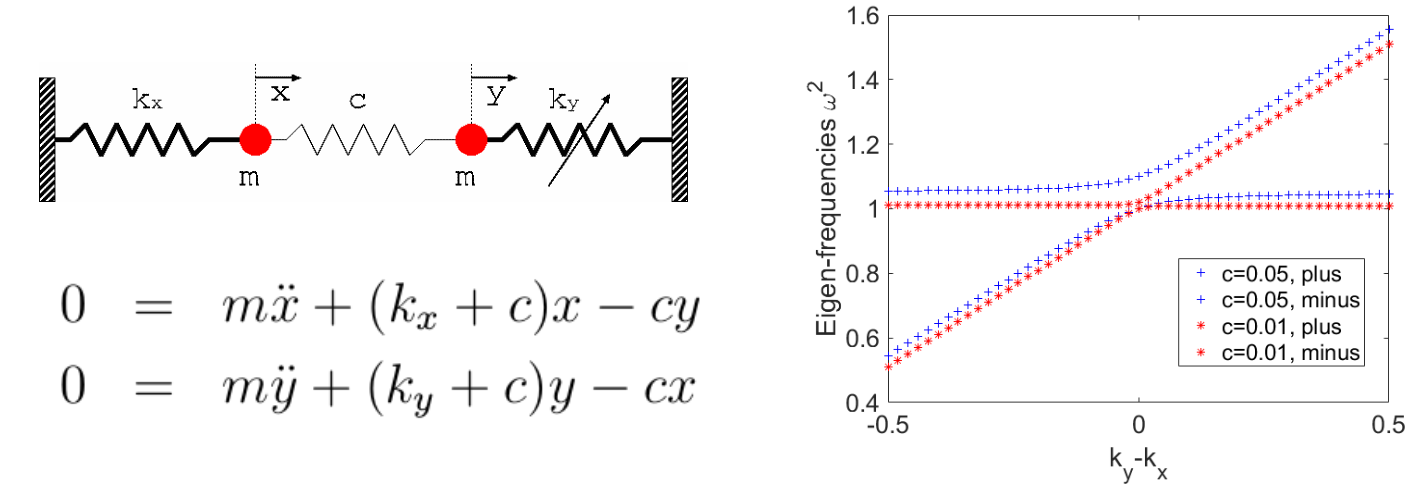}
\end{center}
\caption{\label{fig:coup}Left: mechanical analogy of coupling and the corresponding
  equations of motion. Right: the eigenfrequencies from Equation~\ref{eq:eigtune} for
  two values of the coupling $c$ as a function of $k_x-k_y$}
\end{figure}

Quadrupoles that are accidentally mounted with a roll angle or fields from
solenoids can couple the betatron oscillations in the transverse planes, which
has an influence on the tunes $Q_x$ and $Q_y$. Qualitatively this behavior is
easily understood by considering a mechanical equivalent system of two mass
points connected by springs, as shown on the top left in Figure~\ref{fig:coup}.
The deviations $x$ and $y$ from their equilibrium correspond to the betatron
oscillations amplitudes and the unperturbed tunes correspond to the
eigenfrequencies $Q_x^2 = k_x/m$ and   $Q_y^2 = k_y/m$, while the coupling between
these oscillations originates from the weak coupling constant $c$, the spring
constant that connects the two mass points. It is straightforward to obtain
the equations of motion, shown at the bottom left of Figure~\ref{fig:coup}.
With standard methods to solve coupled linear differential equations, we
find the eigenfrequencies $\omega_{\pm}$---corresponding to the two eigentunes
of the coupled system---to be
\begin{equation}\label{eq:eigtune}
\omega_{\pm}^2=\frac{k_x+k_y+2c}{2m} \pm \sqrt{\left(\frac{k_x-k_y}{2m}\right)^2 + \frac{c^2}{m^2}}\ .
\end{equation}
We see that the root can never vanish, unless the coupling $c$ is zero. The plot
on the right-hand side in Figure~\ref{fig:coup} shows the eigentunes $\omega_{\pm}$
from Equation~\ref{eq:eigtune} plotted as a function of the difference between
$Q_x\sim k_x$ and $Q_y\sim k_y$ for $c=0.05$ and $0.01$. The larger value of $c$
causes the eigentunes to ``repel'' each other more. This observation is exploited
operationally by adjusting upright quadrupoles to make the tunes $Q_x\sim k_x$ and
$Q_y\sim k_y$ as close as possible and then adjusting one or more additional skew
quadrupoles to minimize the tune separation and thereby the coupling~$c.$ This
procedure is commonly referred to as {\em correction of the closest tune.}
%
\subsection{Chromaticity measurement and correction}
In order to measure the chromaticity $Q'$ of a ring we have to change the relative
momentum $\delta=\Delta p/p$ of the beam and observe the corresponding change of
the tune $Q=Q_0+Q'\delta$. As mentioned towards the end of Section~\ref{sec:diperr}
can we change the momentum by changing the RF frequency $f_{RF}$ by $\Delta f_{RF}$
as given by
\begin{equation}
 -\frac{\Delta f_{rf}}{f_{rf}} = \frac{\Delta T}{T} = \eta \delta 
= \left(\alpha -\frac{1}{\gamma^2}\right)\delta
\qquad\mathrm{such\ that}\qquad
\delta=-\frac{1}{\eta} \frac{\Delta f_{rf}}{f_{rf}}\ .
\end{equation}
For a number of different relative momenta  $\delta$ we then measure the tune, for
example, by exciting a betatron oscillation and Fourier-transforming a position signal
from a BPM. From a plot of the measured tune versus $\delta$ we can derive the
chromaticity from a straight-line fit.
\par
The chromaticity is a consequence of the momentum-dependence of the focusing of
quadrupoles and, in order to correct it, we use sextupoles, placed at a location
with non-zero horizontal dispersion $D_x$. The dispersion causes the particles
with relative momentum $\delta$ to have an additional transverse offset $D_x\delta$,
which is equivalent to transversely displacing the sextupole by $d_x=D_x\delta.$ 
Inserting in Equation~\ref{eq:offx} we read off that the sextupole with integrated
strength $k_2L$ produces the field of a momentum dependent quadrupole with
focal length $f_{\delta}$ given by $1/f_{\delta}=k_2LD_x\delta$, which causes 
momentum-dependent tune shifts $\beta/4\pi f_{\delta}$ in the respective planes
\begin{equation}
\Delta Q_x = \frac{k_2L D_x\beta_x}{4\pi}\delta
\qquad\mathrm{and}\qquad
\Delta Q_y = -\frac{k_2L D_x\beta_y}{4\pi}\delta\ .
\end{equation}
Using two sextupoles at a location with different dispersion and beta functions,
we can create a system that allows us to independently control the chromaticities
$Q'_{x,y}=\Delta Q_{x,y}/\delta$ independently
\begin{equation}\label{eq:chromfix}
\left(\begin{array}{c}\Delta Q'_x \\ \Delta Q'_y\end{array}\right)
= \frac{1}{4\pi}
\left(\begin{array}{rr}
D_{1x}\beta_{1x} & D_{2x}\beta_{2x}\\
-D_{1x}\beta_{1y} & -D_{2x}\beta_{2y}
\end{array}\right)
\left(\begin{array}{cc} (k_2L)_1 \\ (k_2L)_2 \end{array}\right)\ .
\end{equation}
Finding the sextupoles excitations $(k_2L)_1$  and $(k_2L)_2$ to change the
two chromaticities by $\Delta Q'_x$ and $\Delta Q'_y$ is now a matter of inverting
the matrix in Equation~\ref{eq:chromfix}.
%
%
\section{Further reading}
Hopefully, reading these pages of introductory material whets your appetite for
more, such as Zimmermann and Minty's book~\cite{MINTY} or the chapter on
operational considerations in the Accelerator Physics Handbook~\cite{HANDBOOK}.
Moreover, several textbooks cover corrections, see for example chapter~6 in~\cite{WOLSKI},
chapter~3 in~\cite{SYLEE}, chapter~7 in~\cite{WIEDEMANN}, and chapter~8 in~\cite{VZA}.
In previous CERN accelerator schools the same topic was covered, see for
example~\cite{JWENN} in the proceedings of the 2009 Diagnostic school,
which also contains contributions on related topics. In general, it is worth
to go poaching in the CAS archives~\cite{CAS} and hunt down the slides of
colleagues who covered similar topics.
%
%
\small
\bibliographystyle{plain}

\begin{thebibliography}{M}
  %
\bibitem{TBD}
  W. Hillert, {\em Transverse linear beam dynamics,} these proceedings.
\bibitem{LBD}
  F. Tecker, {\em Longitudinal beam dynamics in circular machines,} these proceedings.
\bibitem{VZA}
V. Ziemann, {\em Hands-On Accelerator Physics Using MATLAB,}
CRC Press, Baton Rouge, 2019.
\bibitem{MADX}
  MADX project web site: \url{http://mad.web.cern.ch/mad/}
\bibitem{MICADO}
  Y. Marti, B. Autin, {\em Closed orbit correction of A.G. machines using a small
    number of magnets,} CERN-ISR-MA-73-17, 1973.
\bibitem{NR}
  W. Press et al., {\em Numerical Recipes, 2nd ed.,} Cambridge University Press,
  Cambridge, 1992.
\bibitem{BMAG}
  T. Raubenheimer, F-J. Decker, J. Seeman, {\em Beam distribution after filamentation},
  Proceedings of the Particle Accelerator Conference, 1995.
\bibitem{CALIF}
  W. Corbett, M. Lee, V. Ziemann, {\em Model Calibration and Beam Control Systems
    for Storage Rings,} Proceedings of the Particle Accelerator Conf., Washington,
  D.C., 1993.
\bibitem{LOCO}
  J. Safranek, {\em Experimental determination of storage ring optics using orbit
    response measurements,} Nuclear Instruments and Methods A 388 (1997) 27.
\bibitem{MINTY}
  F. Zimmermann, M. Minty, {\em Measurement and Control of Charged Particle Beams,}
  Springer Verlag, Berlin, 2003.
\bibitem{HANDBOOK}
  F. Zimmermann, K. Mess, M. Tigner, {\em Handbook of Accelerator Physics and Engineering,
    2nd ed.}, World Scientific, Singapore, 2013.
\bibitem{WOLSKI}
  A. Wolski, {\em Beam Dynamics,} Imperial College Press, London, 2014.
\bibitem{SYLEE}
  S.Y. Lee, {\em Accelerator Physics, 2nd ed.,} World Scientific, Singapore, 2004.
\bibitem{WIEDEMANN}
  H. Wiedemann, {\em Particle Accelerator Physics I, 2nd ed.,} Springer Verlag,
  Berlin, 2003.
\bibitem{JWENN}
  J. Wenninger, {\em Lattice measurement,} CERN-2009-005 (2009) 361.
\bibitem{CAS}
  \url{https://cas.web.cern.ch/previous-schools}
%
\end{thebibliography}

\end{document}